# PABED – A Tool for Big Education Data Analysis

Samiya Khan
*Department of Computer Science*
*Jamia Millia Islamia*
New Delhi, India
samiyashaukat@yahoo.com

Kashish Ara Shakil
*Department of Computer Science and Engineering*
*Jamia Hamdard*
New Delhi, India
shakilkashish@yahoo.co.in

Mansaf Alam
*Department of Computer Science*
*Jamia Millia Islamia*
New Delhi, India
malam2@jmi.ac.in

*Abstract*—Cloud computing and big data have risen to become the most popular technologies of the modern world. Apparently, the reason behind their immense popularity is their wide range of applicability as far as the areas of interest are concerned. Education and research remain one of the most obvious and befitting application areas. This research paper introduces a big data analytics tool, PABED (Project - Analyzing Big Education Data), for the education sector that makes use of cloud-based technologies. This tool is implemented using Google BigQuery and R programming language and allows comparison of undergraduate enrollment data for different academic years. Although, there are many proposed applications of big data in education, there is a lack of tools that can actualize the concept into practice. PABED is an effort in this direction. The implementation and testing details of the project have been described in this paper. This tool validates the use of cloud computing and big data technologies in education and shall head start development of more sophisticated educational intelligence tools.

## I. INTRODUCTION

Big data is the future of technologies in view of the fact that every other new-generation technology needs to customize itself in accordance with the evolving needs of systems to accommodate for big data scenarios. The biggest positive as well as negative of big data technologies is the range of applications that they are expected to support.

Application areas for big data vary from public welfare sectors like education [1], transportation [15] and healthcare [16], to domains like business intelligence [17] and geospatial data analytics [18]. Considering the fact that this paper largely focuses on applications of big data in education, it is worth mentioning that there are many known and existing application areas.

Some of these include student performance prediction [4], institute management [3], quality management [2], student dropout prediction and online education applications [5] [13]. With this said, these are all proposed applications and research in this field is yet to mature.

It is feasible to migrate established sub-systems like operational management and quality assessment completely to the big data and cloud infrastructures. The tool proposed in this research paper performs validation of this proposal by allowing analysis of big education data using base technologies like Google BigQuery [7] and R programming language [6].

The usage of big data technologies in education is well established. However, tools that can validate this usage are missing. PABED is a project that aims to create education-specific applications using base technologies to solve specific problems and fill this void. Moreover, this project also plans to actualize these prototypes into commercially viable solutions.

Presently, PABED is still under development and allows basic functionality. It allows the user to source data from Google BigQuery, in which tables can be created from Google Drive account. The tool allows the user to compare undergraduate enrollments for two academic years. The total enrollments' value and a line graph to indicate the trend are created as part of the tool.

The existing tool can be extended to include many other features. Some of the proposed features include analysis of the share of different ethnicities in the overall enrollments, gender-wise study of enrollment rate and predictive analysis of how enrollment rate is expected to change in the coming years for different institutions. Moreover, other data features like faculty ratios, organizational infrastructure, budget, expenses and student performance parameters can also be used as base parameters for devising more analytical solutions.

The following sections of this research paper are organized in the below-mentioned manner. Section II elaborates on the use of cloud and big data technologies for development of educational intelligence tools; thereby, commenting on the viability and feasibility of such an application. The methodology and implementation details for the proposed educational intelligence tool called PABED are specified in Sections III and IV. The last section of this research paper makes concluding statements on the high and low points of the proposed tool along with a statement on scope for future research in the same.

## II. BIG DATA IN EDUCATION

Big data technologies when integrated with cloud infrastructures are known to solve many real-world problems. Most of these applications have been classified and termed on the basis of the domain of problem. For instance, business intelligence is the term used to describe big data applications in business [19]. Similarly, applications using big data tools and technologies, developed for the research and education sectors, are called educational intelligence [8] applications.

Recent literature related to the field includes works by Williamson [26] and Klašnja‑Milićević [29], which elaborate on the use of big data in education, giving insights on learning analytics and practical applications of the same. Besides this, Swing [27] and Shah [28] discuss the use of analytical solutions in the field of higher education.

Profile-wise, big education data includes data related to the different actors of the education field. Some of the main actors include students, faculty, non-teaching staff and organization, as a whole. Data related to students, faculty and educational institutes constitute big data for education.

Student data can further be broken down into data related to students' personal profiles, performance scores, attendance and assessment reports for extra-curricular and sports activities. Besides this, once a student passes out and becomes alumni, data related to the same is also part of student data.

On the other hand, educational institute is similar to any other organization with respect to the organizational processes involved. The management and staff of the organization are the two pillars of the organizational setup and any data related to the same is referred to as organizational or faculty data [1]. In addition to this, data related to quality assessment and performance evaluation of the institute is also included.

Apart from the data types already mentioned, another class of data that is an integral part of the modern educational setup is research data. Faculty along with research students generates what can be called 'research data'. The relevance of this data can be attested by the fact that most quality assessment parameters for faculty and educational institutions include research data for evaluation [20]. The data, education and research, is high in volume, includes textual, image and multimedia information, and is generated on a periodic basis, which satisfies the three Vs (volume, variety and velocity) for classification of a dataset as big data [21].

In view of the relevance of data analytics to the education and research sectors, several applications are considered useful. Some of the applications that have well-established reputation in these fields include quality assessment systems for higher education, research management systems, student performance analyzers and business intelligence applications for the education sector. Some prototypes have been proposed related to these application domains. However, research is still in its infancy and no commercially viable solutions are known to exist, which leaves immense scope for future research and development.

### III. METHOODOLOGY

The Project - Analyzing Big Education Data (PABED) aims to create a concept tool for analysis of big education data. As far as technologies are concerned, there are two specific requirements of this tool. Firstly, a data warehousing solution is required, which can be used for data storage. Besides this, the data concerned needs to be accessed, queried, retrieved, manipulated and analyzed. This requires a processing language that can be interfaced with the storage solution as well as the web view presented to the user for making any such requests. In accordance with the requirements of the system, the technologies chosen for the implementation of this educational intelligence tool are R [6] and Google BigQuery [7].

Google BigQuery [7] is deemed appropriate storage solution for this tool for its ability to store huge amounts of data in the cloud. Moreover, the facility is free of cost until the data size goes beyond 10 GB on a monthly basis. Therefore, for data under this limit, this facility is practically free for the user, bringing the solution cost to a bare minimum. Other solutions that can be considered as a replacement for BigQuery are NoSQL databases [22] like MongoDB [23] and Cassandra [24]. However, the complexity of such solutions is extremely high and they must be considered only if the solution requires such high intricacy in design.

The R programming language [6] was found appropriate for the processing needs of the system. Moreover, it is open source and loaded with packages for implementation of different functionalities. Therefore, the complexity of the project is considerably reduced, also decreasing the development time and effort. Other technologies that can be used as a processing solution for the system are Hadoop [10], Hadoop with R [10] and Spark [11]. However, they are expected to increase the complexity and cost of the project. Considering the present status of the project, they shall lead to unnecessary overheads.

R has a dedicated package for enabling an interface with Google BigQuery [7], which allows data retrieval and query processing. Moreover, graphics packages for creating plots and graphs allows visual data analysis. Lastly, one of the key requirements of the system concerned is to develop a web application for user interfacing. R provides a package called Shiny [12] for this purpose.

Fig. 1. User Interface for PABED

Fig. 2. Inputs Section of PABED

The reasons for choosing BigQuery as a storage solution and R programming language as a processing solution for the project can be summarized in Table 1.

TABLE I. SUMMARY OF TECHNOLOGICAL DECISIONS MADE FOR THE PROJECT

| Technology | Reasons for Choosing the Technology |
|---|---|
| Google BigQuery | 1) Free data storage of 10 GB per month is allowed on BigQuery, making it a cost-effective solution. <br> 2) Cloud-based, distributed storage allow easy storage of massive datasets, which is one of the fundamental requirements of the project as far as storage is concerned. <br> 3) Basic data access, retrieval and query abilities of BigQuery provide for the fundamental processing requirements of the project. |
| R Programming Language | 1) Open-source solution, contributing to the cost-effectiveness of the developed tool. <br> 2) Availability of packages and inbuilt functionalities for establishing interface with BigQuery, creation of graphics and development of web views simplifies solution's development and reduces its maturity effort and time. |

IV. IMPLEMENTATION AND TESTING

PABED (Project - Analyzing Big Education Data) is an educational intelligence tool for big education data. Google BiQuery and R programming language are the two main base technologies used for implementation. The user interface has been designed using the Shiny Dashboard package. The use of the same to make user-friendly interfaces has been inspired from BigQuery Visualizer [9]. The user interface for the application can be seen in Fig. 1.

Due to the BigQuery limitation that allows users to create tables using file uploads only if the file size is less than 1 MB, data or individual CSV files have been uploaded to Google Drive and sourced from the same for usage in BigQuery. It is also possible to upload data to Google Cloud Storage. However, the latter option is chargeable and in order to keep the cost minimum, the former approach was chosen.

The Google BigQuery and Google Drive authentication for the application is done on server side using the application key, in the form of a JSON file, downloaded from the concerned BigQuery project service account. The user is required to specify the BigQuery Project ID, Database Name, Academic Year-1 and Academic Year-2. The first two parameters are required for query processing and the user can find them in the BigQuery console for the project concerned.

Presently, the project supports functionality for comparing the total undergraduate enrollments for two academic years. The dataset files are named after the academic year that they belong to. For instance, the dataset for the academic year 1996-97 is named as 1996_97. Therefore, while specifying academic years as inputs the user is expected to give 1996_97 as a value. The dataset has a column that specifies the undergraduate enrollments for all the insitutions in an academic year (UGDS).

The implementation sums the values for the two concerned academic years and creates a line graph to indicate the trend. The resultant values for the two academic years have been displayed on the top-right corner of the line graph. The input values and result and analysis generated for the same can be seen in Fig. 2 and Fig. 3.

The dataset also includes NULL values for some insitutions, for which the data may not be available. These NULL values have been handled by the code. Moreover, since the dataset is sourced from Google Drive and Google BigQuery import for the data was marked to detect schema automatically. In such cases, most fields are automatically detected to contain the datatype STRING. Data conversions that may be required in this regard have also been handled in the implementation.

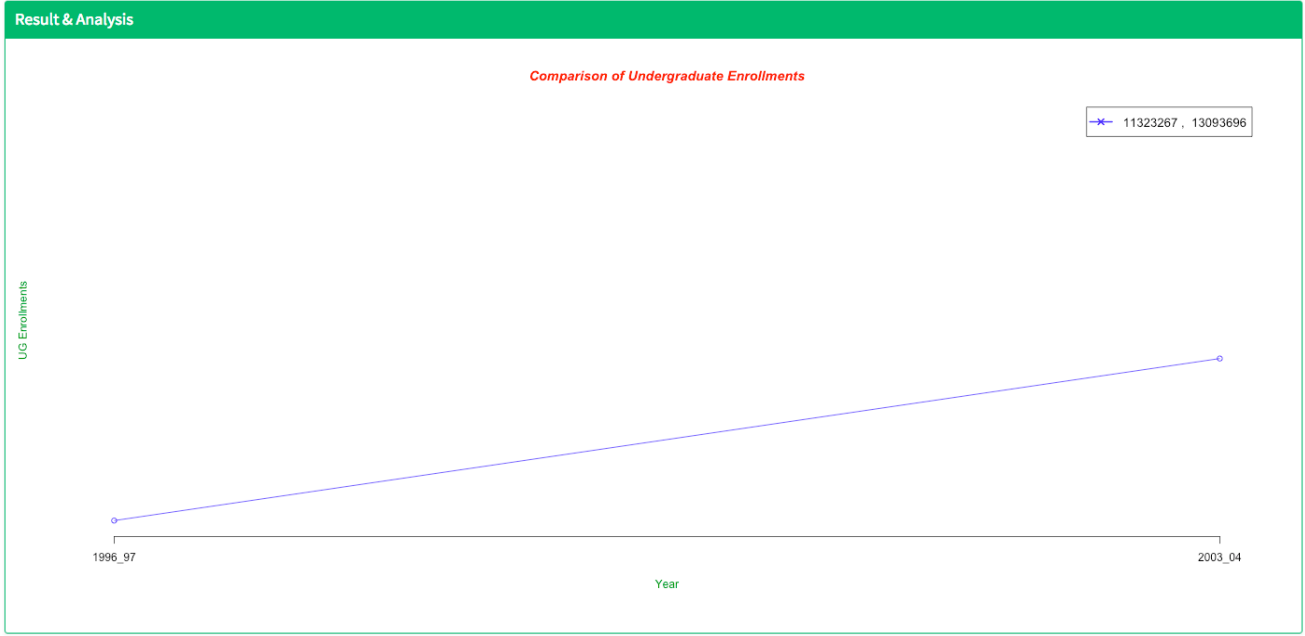

Fig. 3. Line Graph for Comparing the Undergraduate Enrollments for the Academic Years 1996_97 to 2003_04

Considering the fact that the technologies used for the project are well-equipped to handle evolving datasets, the tool is expected to scale up to higher data size demands with ease. The source code for PABED has been uploaded on shiny server for global access and have been made available under the open-source copyright at the web address: https://qmhes.shinyapps.io/qmhes_2/.

The testing of the system has been performed on local server using a dataset [25] taken from US Department of Education and is 2.39 GB in size. The source code for the project has been made available on GitHub under the project name PABED-2. The project is open-source and and can be cloned for personal use. In order to clone the project effectively, the user must follow the instructions given below:

- Upload the dataset to Google Drive and source the dataset tables from Google Drive to Google BigQuery account.
- Download the server.r and ui.r files for the project.
- Generate API key for the Google BigQuery Project and rename it as bigrquery-token.json.
- Place all the three files in the same folder.
- Test the project locally using Rstudio. Be sure to enable Google Drive access before using.
- Lastly, upload the project files to Shiny server account using RSConnect package, for global access.

It is important to mention here that the tool is still under development and thus, it supports minimal functionality. Implementing more sophisticated analytical tasks for the educational field shall be considered in the future. Besides this, the response time of the application is higher than expected. This can be reduced with the use of more sophisticated big data technologies like Hadoop and Spark. Although, the use of these technologies is expected to increase the cost of project, it will improve the capabilities of the application, immensely.

## V. CONCLUSION

This research paper proposes PABED (Project - Analyzing Big Education Data), an educational intelligence tool, for analyzing big education data. It implements determination and comparison of undergraduate enrollments for two academic years with the help of a line graph. The prototype implementation is inspired by that of BigQuery Visualizer [9] and makes use of Google BigQuery [6] and R programming language [7] as the base technologies.

Considering the complexity, cost and design requirements of the project, the chosen technologies were deemed appropriate. The reduced cost and development time or effort for such project makes it a benchmark for development tools that are specific to the requirements of an educational institute. In other words, customization of solutions to suit specific needs shall require lesser budge, time and effort.

PABED attests the use of cloud and big data technologies for education. Existing literature has proposed many applications of big data technologies in the field of education. However, commercially viable tools or even prototype implementations have not been made. PABED is one of the first of its kind effort towards actualizing the use of big data concept in education.

The tool has been designed to fulfil the base requirements of such a system. It is still in the nascent stages of development and can be extended and customized to suit specific requirements of any application. Moreover, depending upon the requirements of the desired system, generic or specific, the use of other sophisticated technologies like Hadoop and Spark shall also be explored in the future.

### STATEMENTS ON OPEN DATA, ETHICS AND CONFLICT OF INTEREST

Project - Analyzing Big Education Data (PABED) is available as an open source project in GitHub (https://github.com/samiyakhan13/PABED-2). A dataset taken from U.S. Department of Education [25] was used to test the application. An electronic version of data will be made available and shared with interested researchers under an agreement for data access (contact: samiyashaukat@yahoo.com).

All participants were informed well about the research objectives, contents and their right to easy withdrawal without reasoning and all gave informed consent. Data were treated anonymously and no personal identifiers were reported. There are no potential conflicts of interest in the work.

ACKNOWLEDGMENT

This work was supported by a grant from "Young Faculty Research Fellowship" under Visvesvaraya PhD Scheme for Electronics and IT, Department of Electronics & Information Technology (DeitY), Ministry of Communications & IT, Government of India.